

\input harvmac.tex

\def\inbar{\,\vrule height1.5ex width.4pt depth0pt}
\def\IB{\relax{\rm I\kern-.18em B}}
\def\IC{\relax\hbox{$\inbar\kern-.3em{\rm C}$}}
\def\IP{\relax{\rm I\kern-.18em P}}
\def\IR{\relax{\rm I\kern-.18em R}}

\def\pa{\partial_{+}}
\def\pab{\partial_{-}}

\Title{\vbox{\baselineskip12pt\hbox{RU-91-54}\hbox{}}}
{\vbox{\centerline{Are Horned Particles the Climax of Hawking
Evaporation?}}}

\centerline{T. Banks$^1$, A. Dabholkar$^2$, M. R. Douglas$^3$, M.
O'Loughlin$^4$}
\baselineskip18pt
\centerline{Dept. of Physics and Astronomy}
\centerline{Rutgers University}
\centerline{Piscataway, NJ 08855-0849}
\smallskip
\baselineskip18pt
\noindent
We investigate the proposal by Callan, Giddings, Harvey and Strominger
(CGHS) that two dimensional quantum fluctuations can eliminate the
singularities and horizons formed by matter collapsing on the
nonsingular extremal black hole of dilaton gravity.  We argue that this
scenario could in principle resolve
all of the paradoxes
connected with Hawking evaporation of black holes.  However, we show
that the generic
solution of the model
of CGHS is singular.  We propose modifications of their model
which may allow the scenario to be realized in a consistent manner.

\vskip1in
\noindent

\footnote{}{$^1$ (banks@physics.rutgers.edu)}
\footnote{}{$^2$ (atish@physics.rutgers.edu)}
\footnote{}{$^3$ (mrd@physics.rutgers.edu)}
\footnote{}{$^4$ (ologhlin@physics.rutgers.edu)}
\Date{January 1992}

\newsec{\bf Introduction}

The apparent paradoxes of the Hawking process of quantum mechanical
decay of a black hole have been a source of confusion and inspiration to
theoretical physicists since Hawking's groundbreaking papers in the mid
seventies \ref\hawk{S.W. Hawking, Comm. Math. Phys. {\bf 43} (1975),199;
Phys. Rev. {\bf D14} (1976), 2460.}.
Hawking proposed that these paradoxes might lead to
a generalization of quantum mechanics in which pure states can evolve
into mixed states. Alternative scenarios in which coherence is
preserved
have often revolved around
the idea that the endpoint of black hole decay would be some sort of
exotic elementary particle.  The apparent
validity of Hawking's analysis
down to black hole masses where quantum gravitational or string
theoretic effects become important provides strong constraints on such a
scenario.

Many authors have expressed discomfort with the apparent information
loss in the decay of a black hole into a stable elementary particle.
This somewhat vague argument can be sharpened in those cases in which
one deals with a theory containing global conserved quantum numbers
such as baryon number.  It is easy to construct models in which
Hawking's analysis is valid until the black hole is so light that it
cannot emit enough ordinary particles to carry its baryon number.
While the force of this argument is somewhat vitiated in the real world
by current theoretical prejudices against the existence of global
conserved quantum numbers, it is still a valid objection to many
theoretical models.  In such models one must postulate an infinite
spectrum of almost degenerate elementary particles representing the
hypothetical endpoints of the evolution of black holes with different
values of baryon number.  This leads to apparent paradoxes with
thermodynamics and with many of the usual rules of quantum field theory,
unless one postulates that the coupling between ordinary particles and
these black hole remnants is extraordinarily weak for some unspecified
reason.

There has for a long time been a widespread feeling that these problems
would not be resolved until the correct short distance theory of gravity
was discovered.  Einstein's gravitational theory certainly needs
modification at small length scales if it is to be made compatible with
quantum mechanics.  It is hoped that the correct modification would also
resolve the paradoxes of Hawking radiation, either at the classical or
quantum mechanical level.  There is, it seems to us, one bit of hope
that the understanding of Hawking radiation will not have to await the
perhaps distant day when a consistent nonperturbative quantum theory of
gravity is discovered.  This is related to the fact that extreme
Reissner Nordstrom (RN) black holes have zero Hawking temperature.  The
geometry of an extreme RN hole of large charge is smooth and varies on
length scales much larger than the Planck scale.  Perturbing this
solution with a small amount of neutral matter one obtains a black hole
of small Hawking temperature.  It is plausible to conjecture that the
evaporation of this black hole proceeds back to the extreme Reissner
Nordstrom solution and then stops.  Since all the physics takes place at
large length scales, we should be able to describe this
process without recourse to the short distance modifications of
Einstein's theory.\foot{This idea has occurred to a large number of
authors over the past fifteen years.  As far as we know, it has been
independently arrived at by one of us (T.B.),
C.Callan, D. Gross, E. Martinec, J. Preskill, L. Susskind, F. Wilczek, and E.
Witten. Probably the list is much longer. The only published versions of
these ideas of which we are aware are\ref\presk{J. Preskill, P. Schwarz,
A. Shapere, S. Trivedi, and
F. Wilczek, Mod. Phys. Lett. A6 (1991) 2353; {\it Black Holes as
Elementary Particles},
C. F. E. Holzhey, F. Wilczek, IASSNS-HEP-91/71 (1991).}}
This idea looks even more attractive
when one considers the extremal black hole of the version of dilaton
gravity that follows from string theory\ref\GHS{G.Gibbons, Nucl. Phys.
{\bf B207},(1982),337; G.Gibbons, K.Maeda,
Nucl. Phys. {\bf B298},(1988),741; D.Garfinkle,
G.Horowitz, A.Strominger, Phys. Rev. {\bf D43} (1991), 3140.}.
When interpreted in
terms of the correct metric variable, this is a completely
nonsingular classical spacetime geometry with no event horizons.
The idea that Hawking evaporation of black holes of large charge might
terminate at this configuration is extremely compelling.

Alas, things are not so simple.  Generic
classical perturbations of the extreme RN solution\foot{Either in
general relativity or stringy dilaton gravity.}, including those
caused by dropping in a bit of extra mass, lead to singularities.  In
the neighborhood of the singularities curvatures become infinite and one
might imagine that the short distance theory becomes relevant.  On the
other hand, it seems a bit implausible that one would have to include
high mass string modes to describe the evolution of a large smooth
classical geometry.
 One might hope that quantum mechanical fluctuations of long
wavelength gravitational fields were sufficient to tame these
singularities.  The question is how to isolate the relevant degrees of
freedom and quantize them in a manner consistent with general
covariance.

Recently, major progress towards a resolution of this problem has been
made in a beautiful paper
by Callan, Giddings, Harvey, and Strominger (CGHS)\ref\CGHS{C.Callan,
S.B.Giddings, J.A.Harvey, A.Strominger, {\it Evanescent Black Holes},
UCSB-TH-91-54, EFI-91-67, PUPT-1294}.\foot{The authors
of this paper present their results in the context of the two
dimensional effective theory described below.  They point out
the connection to higher dimensional black
holes. We will describe their results in the higher dimensional
context.}  They considered the extremal charged black hole derived from
string theory in the work of Gibbons and Maeda, and
Garfinkle {\it et. al.}\GHS .
The solution is spherically symmetric and a generating section of its
spatial geometry is shown in fig. 1.  It is an infinite horn stuck onto
flat space, and has two
asymptotic regions, \lq\lq down the horn '' and , \lq\lq out at
infinity.''
Nonextremal perturbations of this solution have horizons and
singularities, but for small perturbations, the horizons are down the
horn and the singularities are behind them.  It is thus plausible that
we can isolate the degrees of freedom involved in the quantum mechanical
decay of the hole, and the possible elimination of both the singularity
and the horizon, by constructing an effective two dimensional field theory for
massless fields in the horn.  CGHS were thus led to study a model
of two dimensional gravity coupled to a scalar (dilaton) field and
several conformally coupled scalar matter fields, which may be thought
of as dimensionally reduced Ramond-Ramond fields from the type II
superstring. The classical equations of motion of this theory were
exactly soluble and the solutions included the extremal black hole
(whose two dimensional representation is flat space time with a linear
dilaton), as well as solutions in which Ramond Ramond matter collapses
onto the extremal hole to form a singular hole of larger mass.

By adding a large number of RR fields to their Lagrangian, CGHS obtained
a model whose quantum mechanics could be systematically studied in the
${1\over N}$ expansion.  They studied the semiclassical large N
equations of motion and argued that the collapsing solutions of the
classical equations became nonsingular, horizon free solutions of the modified
equations.  The collapsing matter forms a \lq\lq zero energy bound state''
 concentrated in the asymptotic region \lq\lq down the horn''.  There
are an infinite number of such states, and they carry zero ADM energy as
measured by an observer in the other asymptotic region.  They can
however carry global conserved charges.  We will argue in the last
section of this paper that if the CGHS results are correct they suggest
a description of the final state of black hole evaporation in terms of
quantum fields interacting with \lq\lq horned particles '' that carry an
infinite number of degrees of freedom.  The concept of a \lq\lq horned
particle '' or {\it cornucopion}\foot{This name combines a description
of the classical dilaton black hole geometry, with the notion that the
infinite horn of the hole is full of unexpected information.}
is based on the classical geometry of
the extremal dilaton black hole\GHS, and seems to eliminate the paradoxes
associated with scenarios in which black hole evaporation ends in a
stable object.\foot{The suggestion that infinitely degenerate particles
could resolve the unitarity puzzle of black holes was apparently first
made in\ref\ahron{Y.Aharonov, A.Casher, S. Nussinov, Phys. Lett. {\bf
B191}, (1987), 51.}}  From the point of view of an external observer the black
hole shrinks until it occupies a very small region in space.  However,
the infinite horn of the GHS solution is a repository for an infinite
number of quantum states that are degenerate in ADM energy but very
difficult to excite with an external probe.  Following CGHS, we will
argue that this
property of the horn makes it a perfect answer to the conventional
conundrums of black hole evaporation.

It thus seems critically important to understand the semiclassical
equations studied by CGHS and verify that they have the sort of
solutions that the arguments of those authors suggest that they have.
This is what we have attempted to do in this paper.  Unfortunately, we
find instead that the collapsing solutions all have singularities. The
classical Lagrangian used by CGHS becomes degenerate in the region of
dilaton field associated with strong string coupling. When the string
coupling becomes strong, the kinetic energy of the fields vanishes.  The
effect of large N quantum corrections is to add a term (the Liouville
term) to the kinetic energy which does not vanish in this region of
field space.  However, if the conformal gauge Lagrangian is viewed as a
nonlinear sigma model in two dimensional target space, the metric on
this target space is degenerate along a line where the dilaton field
takes on a certain finite value $\phi_0$.  The effect of the Liouville term has
been to move the degeneracy to a finite place in field space.  We are
then able to show that if the dilaton field
attains the value $\phi_0$ at some point in spacetime, a generic
solution of the equations will have singular
curvature at this point.  In particular, this occurs for the solutions
corresponding to collapsing shock
waves, which were studied by CGHS.
This analysis is presented in the next section.

The only solutions of the CGHS equations which pass through $\phi_0$
without encountering a singularity, are singular conformal transforms of
the linear dilaton solution.  We can find such a solution of the
equations for every incoming $f$-wave.  The resulting spacetimes are
flat but geodesically incomplete.  We believe that they represent the
\lq\lq zero energy bound states'' or cornucopions,
envisioned by CGHS as the asymptotic
state of gravitational collapse.  However, their geodesic
incompleteness, and the singular nature of solutions with shock wave
boundary conditions, imply that within the context of the two dimensional
theory, there is no way to connect these solutions on to those obeying
classical collapse boundary conditions.  We speculate that this may be
possible in the context of the higher dimensional theory in which the
CGHS Lagrangian is embedded.

Finally, in the last section of the paper we disregard the difficulties
that we have uncovered and present an effective description of the
interactions of ordinary quantum fields with a single cornucopion.  Our
description is completely compatible with general covariance and with
quantum mechanics.

\newsec{\bf Degenerate Lagrangians With Singular Solutions}
\bigskip
\centerline{\it Review of the work of CGHS}

Quantum gravity coupled to a large number $N$ of
free scalar fields, in the limit $\hbar\rightarrow 0$ with $N\hbar$
fixed, is a candidate for a system
in which the leading order of a controlled perturbative expansion is
a quantum theory of matter in classical geometry.
If we integrate out the matter, the path integral has the form
\eqn\one{
Z = \int [dg]~ exp -{1\over \hbar}(S_G[g] +
 {N\hbar\over 2}\log\det(\Delta[g]+m^2))}
and we have an effective action describing the back reaction of matter
and Hawking radiation on the geometry, which we can hope to treat
classically\ref\largen{E.Tomboulis, Phys. Lett. {\bf B70}, (1977), 361.}.
This theory is problematic in four dimensions; the effective
action for the matter is non-local, and the classical equations of
motion appear ill-defined. The effective action has divergent terms
proportional to invariants formed from the squares of curvatures.
To renormalize the theory one has to admit these terms in the classical
action, and they
lead to instability in classical gravity.

CGHS have exploited the fact that these counterterms do not arise
in two dimensions.  For the classical gravitational action they use
a theory of ``dilaton gravity'' with the action
\eqn\two
{S= { 1 \over 2\pi}\int d^2 x\sqrt{-g}\left[e^{-2\phi}(R+4(\nabla\phi)^2
-4\lambda^2)
-\half(\nabla f)^2\right]}
Here $g$, $\phi$ and $f$ are the metric, dilaton, and matter fields,
respectively, and $\lambda^2$ is a cosmological constant.\foot{As we
intimated in the introduction this form is motivated by the low
energy limit
of string theory.  It is our hope
that a regime exists in which semiclassical, field-theoretic
corrections eliminate a singularity of the classical theory.  If this
hope fails in the theory motivated by classical string theory, all is
not lost.
One could imagine broadening the class of two dimensional Lagrangians
under
consideration, to the more general
theories of \ref\bol{T.Banks, M.O'Loughlin, Nuc. Phys. {\bf B362}
(1991),649.}.
Even starting from string theory, one might be able to justify such an
extension of the class of Lagrangians.  The difficulty that we will
encounter in the CGHS theory occurs at moderate values of the (large $N$
scaled )string coupling.  Stringy quantum corrections might easily give
the effective low-energy action a more complicated dilaton dependence in
this regime.}

In two dimensions, integrating out conformally coupled matter fields
gives the
Polyakov action
\ref\poly{A.M.Polyakov, Phys. Lett. {\bf 103B} (1981), 207.}.
This action consistently describes Hawking radiation and its back reaction
on the metric\ref\hawk{S.M.Christensen, S.A.Fulling, Phys. Rev. {\bf
D15}, (1977), 2088.}.
The full effective action in conformal gauge is then\foot{We note in
passing that
if we rescale the metric by a conformal factor which is a function of
the dilaton, the effect is to produce one of the more general actions
described in \bol .}:
\eqn\three
{S_{\rm eff} = \int d^2 e^{-2\phi}
(-2\pa\pab\rho + 4\pa\phi\pab\phi - \lambda^2 e^{2\rho})
+\kappa \pa\rho\pab\rho - \half\pa f \pab f}
where $\kappa\equiv N\hbar/12$.

The equations of motion and constraint equations in this gauge are:
\eqn\four{
0 = {\delta S\over\delta\phi} =
e^{-2\phi} (4\pa\pab\rho + 8\pa\phi\pab\phi -8\pa\pab\phi
+2\lambda^2 e^{2\rho})}
\eqn\five{
0 = {\delta S\over\delta\rho} = G_{+-} =
2e^{-2\phi} (2\pa\pab\phi - 4\pa\phi\pab\phi
-\lambda^2 e^{2\rho}) - 2\kappa \pa\pab\rho}
\eqn\six{
0 = {\delta S\over\delta g^{++}} = T^f_{++} - G_{++} =
\half(\pa f)^2 +
e^{-2\phi} (4\pa\phi\pa\rho -2\pa^2\phi)
- \kappa ((\pa\rho)^2-\pa^2\rho + t_+(x^+))}
(resp. $T_{--}$).

The non-locality of the Polyakov action before gauge fixing
has a remnant in the constraint equations -- in conformal gauge,
these are determined locally by conservation, but only up to two
arbitrary functions $t_+(x^+)$ and $t_-(x^-)$.
These functions must be determined by boundary conditions, in particular
by the choice of incoming and outgoing quantum states.
Furthermore they are not components of a tensor,
but rather transform in a way which
compensates the anomalous transformation law of the term
$(\pa\rho)^2-\pa^2\rho$
in the Liouville stress tensor. We note that the choice of $t_{\pm}$ is
not a full specification of the quantum state of the system.  The matrix
element of the quantum stress tensor between {\it any} two states of a
conformally invariant field theory is determined by two functions each
of which depends on only one of the light cone coordinates.

The general solution of these equations with $\kappa=0$ was obtained in \CGHS.
With $f=0$, we have a one-parameter family of static solutions,
which in a certain coordinate system are
\eqn\eleven
{\eqalign{e^{-2\phi} &={M\over\lambda}-\lambda^2 x^+ x^-\cr
                     &= e^{-2\rho}\cr}}
For $M=0$ the metric is flat and the dilaton linear in $\log x^+ x^-$,
while $M>0$ are ``black hole'' solutions.
Since the $f$ matter is massless, it propagates freely along null geodesics,
\eqn\fmat
{f=f_+ (x^+) + f_-(x^-)\ .}

The black hole solution can be obtained by sending in matter from
the weak coupling region.
The simplest case to consider is an $f$ shock-wave: as in \CGHS;
take its stress tensor to be
\eqn\twelve
{\half \partial_+ f\partial_+ f= a\delta (x^+ - x^+_0)\ ;}
the solution is then
\eqn\thirteen{ e^{-2\rho} = e^{-2\phi} =-a(x^+ - x^+_0)
\Theta(x^+ - x^+_0) -\lambda^2 x^+ x^- . }
For $x^+<x^+_0$, this is  simply the linear dilaton vacuum while for
$x^+ > x^+_0$ it is identical to a black hole of mass $ax^+_0 \lambda$ after
shifting $x^-$ by $a/\lambda^2$.
The two solutions are
are joined along the $f$-wave.

The $f$ shock wave will suffice to illustrate the properties of the
$\kappa\ne 0$ system.
In fact, by rescaling $\lambda$ and the coordinates, we can transform
an $f$ wave of any finite extent to have arbitrarily small extent,
so a smooth, non-singular solution must have a shock wave limit
with non-singular metric and dilaton.

Ignoring back reaction, one finds Hawking radiation, with energy flux
approaching a non-zero constant at late times.  Clearly the back
reaction must drastically modify this, and it becomes important when
$e^{-2\phi} \approx \kappa$, comfortably before the
radiated energy would have equaled the incoming energy.

The primary questions at this point are: is the solution of these
equations of motion with initial data along the shock wave
non-singular, and if so, what does it look like at late times?
An attractive picture is given by \CGHS.
If in the system at late times,
the dilaton behaves as it did in the vacuum,
quantum corrections will be negligible in the weak coupling asymptotic region,
while the strong coupling region will be controlled by the Liouville
action.  This allows only flat space as a vacuum solution, and shock wave
perturbations are described by patching together flat solutions.
This region can be thought of as a ``remnant object'' which contains
the information and conserved charges of the incoming $f$ wave.
\bigskip
\centerline{\it Singular Shock Waves and Incomplete Cornucopions}

All this assumes the existence and non-singular nature of the solution.
A glance at our effective Lagrangian reveals a potential problem with
the CGHS scenario.  In conformal gauge the model resembles a nonlinear
$\sigma -$ model Lagrangian with two dimensional target space. The
target space fields are the Liouville field $\rho$ and the dilaton $\phi$.
The metric on the target space is defined by the kinetic term in the
Lagrangian, and has
determinant $4e^{-2\phi}(\kappa - e^{-2\phi})$.
For unitary matter $\kappa>0$ and this degenerates at a physical value
of the dilaton, $e^{-2\phi} = \kappa$.
This degeneration of the target space metric can cause space time
singularities in solutions of the field equations.  Consider for example
trying to do weak field perturbation theory in
the amplitude of the $f$ wave in order to solve the full field equations
with $\kappa\neq 0$.
This requires us to invert the kinetic term in the linear dilaton
background,
and its degeneration causes this perturbation theory to be divergent
\CGHS .

We can avoid the use of perturbation theory by attempting to solve the
equations of motion directly.
The shock wave problem has vacuum boundary conditions on $I^-_R$,
and boundary conditions on the shock wave determined by continuity.
We will set up a power series expansion in $x^+$ for the
solution, and find that the first derivative of the metric and dilaton
diverges on the shock wave.

It is convenient to use coordinates $\sigma^\pm$ in which the vacuum
(linear dilaton)
metric is $\eta_{ab}$, and
\eqn\dil{\phi_0 = {\lambda\over 2}(\sigma^--\sigma^+).}
(Weak coupling is
$e^{-2\phi}\rightarrow\infty$, $\sigma^+\rightarrow\infty$,
$\sigma^-\rightarrow-\infty$).
In these coordinates $t_+=t_-=0$.

Let the $f$ shock wave be at $\sigma_+=0$ with
$T^f_{++}=a\delta(\sigma_+)$,
and take $\phi=\phi_0+\theta(\sigma^+)\tilde\phi$,
$\rho=\theta(\sigma^+)\tilde\rho$, then
for $\kappa=0$ we have
\eqn\tila{
\tilde\phi=-\half\log(1-{a\over\lambda}e^{\lambda\sigma^-}
(1-e^{-\lambda\sigma^+})),}
\eqn\tilb{\tilde\rho=\tilde\phi.}

To set the boundary conditions for the equations of motion just after
the shock wave, we require the metric and dilaton to be continuous and
have finite derivatives, allowing us to write
\eqn\expa{\phi=\phi_0 + \phi_1(\sigma^-)\sigma^+ + \ldots,}
\eqn\expb{\rho= \rho_1(\sigma^-)\sigma^+ + \ldots.}

The leading term in $\sigma^+$ of the equations of motion are then
($f'$ is $df/d\sigma^-$)
\eqn\eqa{
0 = T_{+-} = e^{-2\phi}(2\phi_1' - 2\lambda\phi_1) - \kappa \rho_1'}
\eqn\eqb{
0 = \delta S/\delta\phi = \rho_1' + \lambda \phi_1 - 2 \phi_1' }
and
\eqn\eqc{\eqalign{
a =& \int_{-\epsilon}^\epsilon d\sigma^+ G_{++} \cr
=& \int d\sigma^+ (2e^{2\rho-2\phi} \pa e^{-2\rho} \pa\phi
- \kappa (\pa^2\rho-(\pa\rho)^2)) \cr
=& 2e^{-2\phi}\phi_1 - \kappa\rho_1.}}

\def\lsm{\lambda\sigma^-}

For $\kappa=0$ we have $\phi_1=a e^{2\phi}/2$, $\rho_1=\phi_1+{\rm
const}$
which agrees with the exact solution.

This analysis suffices to give the boundary conditions for the equations
of motion; continuing the series expansion should give a finite radius
of convergence $R_c$ for any finite $\sigma^-$.
However, this radius could go to zero for large $\sigma^-$, or for large
values of the fields.
In particular, we assumed that $\sigma^+$ was small compared to any
function of $\sigma^-$.
This means that we cannot see the singularity of the original classical
solution at
$a \sigma^+ e^{\lambda\sigma^-} = 1$.
Furthermore we assume $\rho_1\sigma^+$ and $\phi_1\sigma^+$ are small --
if this breaks down, $R_c\rightarrow 0$.

For $\kappa\ne 0$, combine (\eqa) and (\eqb) to get
\eqn\eqab{
\phi_1' = {\lambda\over 2}\phi_1
\left({2 e^{-2\phi}-\kappa\over e^{-2\phi}-\kappa}\right)}
with solution
\eqn\sol{\eqalign{
\phi_1 =& {a\over 2}e^{\lsm}
 \left(1 - \kappa e^{\lsm}\right)^{-1/2};\cr
\rho_1 =& {a\over \kappa}
 \left( (1 - \kappa e^{\lsm})^{-1/2} - 1 \right).}}

As $e^{-\lsm}\rightarrow\kappa$, these blow up, and the curvature
$R= -2\pa\pab\rho \sim \rho_1'$ blows up as well.

The radius of convergence of the series expansion $R_c$ goes to zero at this
point, since
$\rho_1\sigma^+$ and $\phi_1\sigma^+$ are not small.

Thus, an attempt to expand the solution just above the shock wave runs
into a singularity at the spacetime point where the shockwave meets the
timelike line where $1 = \kappa e^{2\phi (\sigma )} = \kappa e^{2\phi_0}$
in the linear dilaton vacuum.  Note that by varying $x_0^+$ for fixed
$a$ and $\kappa$, we can find solutions for which this new singularity
is either behind or in front of the horizon of the classical black hole.
To show that this singularity is not just a failure of the power series in
$\sigma^+$ we study the general solution around a point where $\phi =
\phi_0$.  We attempt to solve the equation near this point by first
writing $\phi = \phi_0 +
\Delta $ .  By taking linear combinations of the trace and dilaton
equations we obtain
\eqn\trace{(e^{2\Delta} - 2)\pa\pab\rho = - 2 \pa\pab \Delta}
\eqn\dil{(e^{2\Delta} - 1)\pa\pab\rho = {3\over 2} (4\pa\Delta\pab\Delta
+ \lambda^2 e^{2\rho})}
Both of these equations exhibit the potential for curvature
singularities, but the second is much more dangerous.  The right hand
side depends only on first derivatives of the field.  If a solution
takes on the value $\Delta = 0$ anywhere in spacetime, the right hand
side of \dil must vanish at that point in order to avoid a
curvature singularity.  But the first derivatives of a solution at a
given point generally change if we change the initial data or the
driving term in the equation.  Thus generic solutions that take on the
value $\Delta = 0$ will have curvature singularities.

Solutions which avoid being singular when $\Delta$ vanishes must satisfy
the restricted equations obtained by setting $\Delta = 0$ above:
It follows that a linear combination of the Liouville field and
dilaton must satisfy the massless wave equation,
\eqn\combwaveq{\pa\pab (2\Delta - \rho ) = 0}
and that
\eqn\othercomb{4\pa\Delta\pab\Delta + \lambda^2 e^{2\rho} = 0}
Thus
\eqn\dileqconfplus{\Delta = {\rho\over 2} + \Delta_+ (\sigma^+ ) +
\Delta_- (\sigma^- )}
Plugging the latter equation into the constraint equations we obtain
\eqn\constrone{\half (\pa f)^2 + \kappa [ (\pa \rho )^2 + 4
\pa\Delta_+\pa\rho - 2\pa^2 \Delta_+ - t_+ ] = 0}
\eqn\constrtwo{ (\pab \rho )^2 + 4
\pab\Delta_-\pab\rho - 2\pab^2 \Delta_- - t_- = 0 }
Note that these are algebraic equations for the derivatives of the
Liouville field with respect to each of the light cone coordinates, with
coefficients which depend only on the corresponding coordinate.  Thus
the Liouville field must be the sum of two functions, each of which
depends only on one of the light cone coordinates.  We see that
near the degeneracy point of the Lagrangian the metric must be
flat, which also
implies that the dilaton must satisfy the massless wave equation.
The two pairs of functions $\Delta_{\pm}$ and $\rho_{\pm}$ must also
satisfy the constraint \othercomb .

In fact, there are exact solutions of the full nonlinear equations satisfying
the constraint that the curvature is zero and the dilaton a massless
free field.    For these solutions, the dilaton and Liouville
field are both solutions of the free wave equation ($\phi = \phi_+
(\sigma^+ ) + \phi_- (\sigma^- )$ and $\rho =  \rho_+
(\sigma^+ ) + \rho_- (\sigma^- )$) satisfying the additional constraints
\eqn\Schrone{\partial_{\pm}^2 e^{-\rho_{\pm}} = - \tau_{\pm} e^{-\rho_{\pm}}}
\eqn\Schrtwo{\tau_{\pm} = t_{\pm} - {1\over 2\kappa}(\partial_{\pm}
f)^2}
\eqn\dilone{\partial_{\pm}\phi_{\pm} = \pm {\lambda\over 2} e^{2\rho_{\pm}}}
 The metric of these solutions is of course
locally flat, but our coordinates cover only a finite region of
Minkowski space.  To see this, note that for a normalizable
left moving $f$-wave the
equation \Schrone has the form of a zero energy Schrodinger equation
for $e^{-\rho}$ with a potential which falls to zero at infinity.  The
asymptotic solutions of this equation are $e^{-\rho}\rightarrow A_{\pm}
\sigma^+ + B_{\pm}$ as $\sigma^+ \rightarrow \pm\infty$.
If we choose a flat metric
in a Minkowski coordinate system in the region that the $f$-wave has not
yet reached, then $A_- = 0$.  For a generic potential (generic $f$-wave
profile) the solution of the zero energy Schrodinger equation will have
nonzero $A_+$.  The coordinate in which the metric is Minkowski is
related to $\sigma^+$, for large positive $\sigma^+$, by
\eqn\coordtran{{dy^+ \over d\sigma^+} = e^{2\rho_+} \rightarrow ({1\over A_+
\sigma^+})^2}
and it is easy to see that $\sigma^+ = \infty$ is a finite point in $y^+$
so that the original spacetime is geodesically incomplete.

It is plausible that these solutions represent the asymptotic \lq\lq
zero energy bound states'' of CGHS, although their geodesic
incompleteness is somewhat puzzling.  What we have shown above is that
any nonsingular solution must locally resemble one of these zero energy
bound states near the point where the Lagrangian degenerates.
Furthermore, the initial conditions defined by the shock wave solutions
of CGHS do not evolve into something that satisfies this condition, but
instead they define a solution which develops a singularity.  This is
indicated by our initial power series analysis.  The existence of
singularities in the generic solution of these equations is an assurance
that the singularity we saw there is not an artifact of the breakdown of
the series.  The power series solution does not satisfy the crucial
condition \dilone . While this argument is not completely rigorous, we
feel confident that it is correct.
\bigskip
\centerline{\it Some Modified Proposals}

Thus, in the two dimensional field theory defined by CGHS, there is no way to
connect the weak coupling region with the strongly coupled Liouville
region without crossing the singularity at $1 = \kappa e^{2\phi} $.
One may attempt
to avoid this difficulty by appealing to the higher dimensional theory
of which the CGHS Lagrangian is a two dimensional approximation.
Indeed, in the higher dimensional black hole solutions, the dilaton
asymptotes to a constant value at infinity.  Thus, at least for a
sufficiently large value of the asymptotic coupling one might hope to
stay below the degeneracy in field space over all of space time.
It is somewhat peculiar to have to require that a coupling be
sufficiently large in order for a semiclassical calculation to make
sense.  Note however that it is the scaled large N coupling $g^2 N$
which is required to be larger than some finite number.  Quantum
fluctuations of fields besides the $f's$ are still suppressed.

In order to write a classical two dimensional field theory which will
have the correct asymptotics of the full four dimensional black hole, we
need only include one extra scalar field in the theory.  This is the two
dimensional representation of a fluctuating radius of the two sphere.
The four dimensional line element is
\eqn\fourdline{ds^2 = g_{\mu\nu}d\sigma^{\mu} d\sigma^{\nu} -
e^{2\Sigma (\sigma)} d\Omega^2}
Here $g_{\mu\nu}$ is the two dimensional metric in the radius-time plane
and $\Omega$ is the three dimensional solid angle.
The two dimensional Lagrangian for these variables is
\eqn\twodlag{\sqrt{-g} e^{-2\phi} \bigl{[} e^{2\Sigma} [R +
2(\partial\Sigma )^2 + 4 (\partial\phi )^2 - 8\partial\phi\partial\Sigma
] - 2 + Q^2 e^{-2\Sigma}\bigr{]}}
where $Q$ is the magnetic charge of the black hole.

Unfortunately, we cannot simply couple this theory to N scalar matter
fields if we want to obtain a sensible system in which ordinary
classical dilaton gravity is recovered at asymptotically large distances
from the center of the black hole.  If the scalar fields are massless
all over the four dimensional spacetime their nonlocal effective action
will be important even at infinity and they will modify the classical
solutions there.  The only way to prevent this would be to take the
scaled coupling $g^2 N$ at infinity very small, in which case we would
encounter the same singularity that we found before.\foot{It is easy to
verify that adding the $\Sigma$ field to the large N action does not
remove the degeneracy.}  Instead we must find a model in which the
quantum fluctuations of the matter fields are naturally suppressed at
large four dimensional distances, and are only important inside the horn
of the black hole.

We have found one such model, though there is no reason to expect it to
be unique.  Add to the dilaton-graviton Lagrangian of low energy string
theory an $SU(2)$ gauge theory with Higgs field in the adjoint
representation, via the prescription ${\cal L}_{flat} \rightarrow e^{- 2\phi}
{\cal L}_{min}$ where ${\cal L}_{flat}$ is the flat space Lagrangian of
the gauge-Higgs system and ${\cal L}_{min}$ is the same Lagrangian
minimally coupled to the stringy metric.  In addition, add $N$ fermions
in the doublet representation of $SU(2)$ with Lagrangian
\eqn\fermlag{{\bar\psi} i e^{\mu}_m \gamma^m D_{\mu} - H^a \tau^a \psi}
The fermions are minimally coupled to the stringy metric.  If they really
corresponded to low energy string excitations, they would also have a
characteristic derivative coupling to the dilaton.  We do not know
whether this affects the discussion of zero modes that we will give
below. Since at the moment we are only interested in finding one model
that works, we can omit the stringy dilaton-fermion coupling if necessary.

In flat space, the Dirac equation in an 't Hooft Polyakov monopole
background has exact normalizable zero energy modes which fall
exponentially with the distance away from the monopole.
As a consequence of the infinite length of the black hole horn, the
Dirac equation in the extremal monopole background has not just isolated zero
modes, but a continuum starting at zero corresponding to modifications
of the zero mode wave function by multiplication with a two dimensional
plane wave in the infinite horn.  The effective action for the quantum
fluctuations of these modes is just $N$ copies of the massless two
dimensional Dirac equation.  By bosonization, we recover the $f$ fields
of CGHS.  Note however that all of these modes are localized in four
dimensional spacetime within a fermion Compton wavelength of the black
hole.  Thus, virtual fermions in these states do not interact with
fields far from the hole, and the contribution to the effective action
from quantum fluctuations of these modes
will not make a substantial change in the equations of motion far from
the black hole.  Contributions to the effective action from massive
fermion scattering states will have the usual derivative expansion in
inverse powers of the fermion mass and will have a similarly innocuous
effect on the large distance equations of motion.

Imagine now a spherical shell of fermion matter collapsing inward from
infinity onto the extremal dilaton black hole of this model.
Take the ADM energy of this shell to be much smaller than the black hole
mass, which is much larger than the Planck scale (they are all of order
$N$).  This will ensure that the formation of a horizon in the classical
evolution occurs only when the collapsing matter has penetrated deep
into the horn of the black hole.
 Take the coupling at infinity to be such that the coupling at
the position of the black hole throat (in the static extremal solution)
is larger than the critical value $e^{-2\phi_0} = \kappa$.
In the four dimensional region far from the throat of the black hole
the evolution of this system will follow the classical equations of
motion with the stress tensor appropriate to the spherical shell of
matter.  As the collapsing shell approaches the throat of the black
hole, our neglect of fermionic quantum corrections to the equations of
motion is no longer justified, and there is a complicated region right
around the throat which we are unable to analyze.  It seems plausible to
assume that an observer some distance \lq\lq down the horn'' from
the throat of the black hole, will see most of the collapsing
matter coming at him in the form of massless $f$-waves.

The problem of determining the further evolution of the system inside
the horn is now very close to that posed by CGHS.  The difference is
that we now envision the two dimensional world of the effective theory
to be only semi-infinite.  The right half of the world (the side which
points towards the throat and the four dimensional world outside it)
terminates at a point where the coupling (varying according to the
linear dilaton solution) is greater than the critical value.  To the
right of this point the two dimensional effective theory is not valid.
A plausible set of boundary conditions for an $f$-shockwave is shown in
fig. 2.  Boundary conditions must be imposed along a segment of the shock
stretching from left future null infinity to the intersection of the
shock with the timelike line $T$ where $e^{- 2\phi} = c < \kappa$, as well
as along this timelike line itself.  In the shaded region we have the
linear dilaton vacuum, and this defines the boundary conditions on the
shock.  We do not understand the appropriate boundary
conditions to put on the timelike line $T$.  In some sense they should
be static, {\it i.e.} all fields should be constant along this line,
because we are attempting to describe
the situation at times after the infalling
matter has passed the point whose world line is T.  Our uncertainty
about these boundary conditions has prevented us from coming to definite
conclusions about the evolution of the system.  The analyses that we are
able to do (such as power series expansion of the solution in the
vicinity of the shock) reveal no singularities, but we have no complete
argument that the dilaton does not pass through $\phi_0$ at some point
in the interior of the unshaded region in fig. 2, thereby causing a
spacetime singularity.  If it were within our power, we would attempt to
demonstrate that these solutions had a nonsingular time evolution which
asymptotically approached the nonsingular solutions of the CGHS
equations which we have identified with cornucopions.  Since the
latter solutions would no longer be thought of as describing a fully
infinite two dimensional world, their geodesic incompleteness would no
longer be a problem.

Our modification of the argument of CGHS depends crucially on re-embedding
their two dimensional problem in the four dimensional problem from which
it came.  Another possible avenue of research is to study modifications
of the two dimensional Lagrangian, within the general class of
Lagrangians described in\bol . In the context of string theory we
may think of this generalization as a way of incorporating higher orders
in the string loop expansion.  More generally we can think of these
corrections as coming from integrating out modes of the higher
dimensional theory which are effectively massive in the horn of the
black hole.  A typical example would simply be higher partial wave modes
of the fermions in the model we described above.

We do not know if a Lagrangian can be found which satisfies all relevant
criteria.  It must reduce to the CGHS Lagrangian for very weak coupling.
It must have static solutions corresponding to a one
parameter family of black holes, with the extremal member of the family
singularity and horizon free.  When the theory is modified by
appending the Liouville action to it,
collapsing solutions should all be singularity free.  The virtue of this
approach when compared to our higher dimensional scenario is that there
is no necessity to assume that the asymptotic coupling in four
dimensions is greater than some critical value.\foot{In superstring
theory the asymptotic coupling is a boundary condition that theorists
hope will be fixed by nonperturbative quantum dynamics.  Dine and
Seiberg\ref\dine{M.Dine, N.Seiberg, Phys. Lett. {\bf 162B}, (1985),
299.} have argued that it
cannot be fixed at a very small
value.  The possibility of fixing it depends on competition between
terms in the effective potential which are {\it a priori} of different
orders in the coupling.  It is perhaps too bizarre to speculate that the
nontrivial lower bound on the coupling that is required for nonsingular
evolution of black holes in our scenario might be related to the
superstringy mechanism which fixes the value of the coupling.} It
clearly deserves further investigation.

To summarize the rather unfortunate situation: CGHS have proposed an
extremely attractive quantum endpoint to the Hawking evaporation
process. Their mechanism does not appear to work in the original model
which they proposed. Instead, we have demonstrated that generic
solutions of their equations, including solutions that correspond to
matter collapsing on the extremal black hole, are singular.
We have proposed modifications of their model in
which respectively, the higher dimensional aspects of the problem (most
importantly the fact that in the four dimensional extremal black hole
solution the coupling is bounded from below) or higher order quantum
corrections to the two dimensional action, remove the singularities.
There is a plausible argument that the first
of these mechanisms leads to singularity free evolution if the
asymptotic value of the coupling is greater than some critical value.
However, our inability to solve the full four dimensional problem with
quantum corrections, or to understand the precise boundary conditions
that it imposes on the effective two dimensional theory, makes this
argument less than convincing.

\newsec{\bf Jumping to Conclusions}

Let us suppose that the difficulties that we have discovered in the
approach of CGHS can be overcome, and that a sensible quantum evolution
of an extremal black hole perturbed by infalling matter can be achieved
in some version of dilaton gravity. Suppose further that an analysis of
uncharged black holes in the quantum regime comes to similar
conclusions.  What can we deduce about the description of the resulting
stable quantum states, and how does this description avoid the problems
with evolution of a black hole into a stable quantum particle that we
reviewed in the introduction?  The first general statement that we can
make is that the extremal black hole will be a new kind of particle
state, which we have termed a \lq\lq horned particle'' or cornucopion.

Indeed, to an observer in the asymptotically flat region, the extremal
black hole with small charge looks like an essentially pointlike object.
Nonetheless, in the context of quantum field theory, it is obvious that
this particle has an infinite number of states.  The semiclassical
geometry is infinitely extended in the direction \lq\lq inside the
hole'', and quantum fields propagating in such a background will have an
infinite number of states whose wave functions fall off exponentially as
we move away from the location of the hole into the asymptotically flat
region. Each of these states is thus another localized particle and we
will argue below that all
of these particles will have very close to the same ADM energy.

The existence of horned particles immediately resolves many of the
apparent paradoxes of Hawking's picture of black hole evaporation.  Even
if we believe the slightly fuzzy arguments about the impossibility of
extracting the information lost to the black hole in the last few Planck
times of its evaporation, we do not have to give up quantum mechanics,
nor even resort to Dyson's hypothesis that black hole evaporation ends
with the creation of a baby universe\ref\dyson{F. Dyson, Institute for
Advanced Study preprint, 1976, {\it unpublished}}.  The information lost in the
formation of the black hole may be viewed as the information contained
in the state of the horned particle.  Any global quantum numbers that the
theory might possess are likewise carried by horned particles.  Given the
semiclassical picture of a horned particle as a geometry with a
semi-infinite tube sticking out of ordinary spacetime, the existence of
almost degenerate states with arbitrarily large values of baryon number
is no longer surprising. It is an obvious consequence of the fact that
the field modes in the horn carry baryon number, and that states
concentrated in the horn do not contribute to ADM energy.  \foot{We will
see below that there should be some degree of nondegeneracy of the ADM
energies of different states confined to the horn.}

The question that we must address is whether it is possible to find a
description of such horned particles which is consistent with quantum
mechanics, general covariance, and ordinary experience.  The crucial
issue is to understand why it is very difficult to produce such
particles in ordinary circumstances, or to excite their infinite number
of internal states.  For example, the existence of an
infinite number of almost degenerate particle states makes even the
microcanonical partition function of statistical mechanics ill defined.
Thus one must understand why it is impossible to produce cornucopions
in ordinary circumstances even when the energy to create them is
available.  Systems of ordinary particles, even at energies much higher
than the cornucopion mass, must not be able to come into equilibrium
with most internal cornucopion states.

We approach this issue by first studying the path integral description
of a single cornucopion interacting with quantum fields of wavelength
much larger than the size of the cornucopion throat.  The particle
follows a world line $x^{\mu}(\tau )$ as a function of its proper time
$\tau$.  We assume that fluctuations of its internal geometry are small,
and choose coordinates $\tau$ and $\eta$ to describe its internal
structure. $\tau$, as mentioned above, is the proper time for the motion
of the particle in the external spacetime, and $\eta$ is the proper
distance from some fiducial point at the mouth of the classical
geometry.  Quantum fields propagating on the background geometry may be
expanded in a complete set of functions of $\eta$.  Internal fields will
be associated with elements of this complete set with support inside the
horn while external fields may be defined as those associated with
functions which vanish at values of $\eta$ larger than some small fixed
constant $d$.  There is of course some arbitrariness in this
distinction, and it is not a useful one for a description of processes
going on near the mouth of the horn.  However, for purposes of our
effective field theory, this arbitrariness is immaterial.  Internal
fields $\phi_i (\eta, \tau)$ will carry a label $i$ which describes,
among other things their properties under rotation of the underlying
spherically symmetric classical geometry; they will be tensor spherical
harmonics.  External fields $\chi_A (x)$will be taken to depend only on
the external
spacetime coordinates $x^{\mu}$.  The label $A$ will describe, among
other things, the particular function of compact support through which
they probe the internal geometry.  Thus, the interaction Lagrangian
between external fields and the internal structure of the cornucopion
takes the form:
    \eqn\Lint{{\cal L}_{int} = \int d\eta d\tau {\cal O}_i (\eta ,\tau )
f_A^i (\eta ) F_M [x^{\mu}(\tau )] O^{A;M}(x^{\mu}(\tau ))\sqrt{-g
(x(\tau )}}
Here, $M$ is an external spacetime tensor index, and $F_M$ is a function
of $x^{\mu}$ and its proper time derivatives.  The functions $f_A^i$
have compact support concentrated near the mouth of the horn. ${\cal O}_i
$ and $O^{A:M}$ are complete sets of composite operators constructed out
of the internal and external fields $\phi_i$ and $\chi_A$ respectively.
Note that this expression is manifestly invariant under diffeomorphisms
of the external spacetime, while it picks out a special coordinate
system on the part of space time swept out by the world history of the
cornucopion.

This interaction Lagrangian must be supplemented by terms in the action
describing the
internal dynamics of the horn, ${\cal S}_I [\phi_i (\eta ,\tau )]$, and
the dynamics of the external fields ${\cal S}_E [\chi_A (x)]$.  In
principal all three of these terms can be computed from the structure of
${\cal S}_E [\chi_A (x)]$ alone by plugging in the decomposition of
fields described above.  For many choices of this fundamental action,
the system will have a set of local and global conservation laws, and
for all systems we have conservation laws associated with the stress
energy tensor in the asymptotically flat external spacetime.  To study
these it is convenient to separate out a set of collective coordinates
from the internal variables of the horn.  We have already done this with
the position coordinate $x^{\mu} (\tau )$.

Given the Lagrangian that we have written, it is very easy to understand
how the system it describes could have an infinite number of states that
are degenerate, or almost degenerate, in ADM energy.
The ADM energy functional
depends only on the values of the fields $\chi_A$ in the asymptotically
flat region of spacetime.  Throught the interaction term ${\cal
L}_{int}$, the internal fields $\phi_i$ can act as sources for the
$\chi_A$ and thereby influence the asymptotic ADM energy.  However,
there are many classical configurations of the $\chi_A$ for which the
interaction Lagrangian is the same.  Any change in $\chi_A$ which has
support only for $\eta \geq d$ does not change ${\cal L}_{int}$.  More
realistically, in quantum mechanics we can imagine that the wave
functionals of many states will be concentrated on field configurations
which vanish rapidly away from some finite place in the horn.  These
states will have an
exponentially small leakage into the region $\eta \leq d$ where the
interaction Lagrangian has its support.
 For example, such a state would be created
by a wave packet of particles travelling down the horn.
Excitations of the horn in these states will cause very tiny changes in
the ADM energy.  In the asymptotic limit in which the wave packet
approaches the end of the horn, it will make no change at all.  These
are the zero energy bound states of CGHS.

The Lagrangian we have written above appears to describe the
hypothetical CGHS endpoint for black hole evaporation in a manner that
is completely consistent with both quantum mechanics and general
relativity.  It describes a system with the ability to absorb the \lq\lq
information '' that has gone down the black hole during earlier stages
of its evolution. In particular, although the cornucopion looks
pointlike to an external observer, it has an infinite number of
degenerate
states and can carry any value of global conserved quantum numbers.
Paradoxes involving thermodynamics are avoided because the external
observer finds it very difficult to excite most of the states of the
horn.  These states are not difficult to excite because they have high
energy, but because they are far away \lq\lq down the horn'' from the
external observer.
Under normal circumstances they would never come into equilibrium
with a gas of particles in the external spacetime, and will not lead to
paradoxical infinities in statistical mechanics or quantum field
theory.\foot{
This argument
refers to equilibration of a system of external fields with an already
existing cornucopion.
The problem of
pair creation of horned particles at sufficiently high energies and
temperatures is also relevant to these thermodynamic questions.  It
is similar to questions about soliton creation in ordinary
field theory.  There, we would expect the amplitudes for these processes to
be very tiny because the coupling is weak.  For the case of cornucopions
there might be additional suppression because of the infinite
volume of space within the horn.  It is even possible that this sort of
infinitely extended object avoids the contraints of crossing symmetry
altogether, and that the amplitude for quantum pair creation of horned
particles is strictly zero, even though scattering amplitudes off
already existing particles are finite.
It seems reasonable that the most efficient
way to create them is through classical gravitational collapse.}

  It is
reasonable to suppose that a unitary S-matrix exists connecting states
on both asymptotic regions of the spacetime.  In most practical
circumstances, the external observer would not be in a position to
obtain information about particles scattered down the horn, and would
have to content himself with measuring inclusive cross sections and
constructing a density matrix.  However, there is no issue of principle
here. In principle one could imagine constructing tiny detectors and
anchoring them at some position deep in the hole\foot{Remember that in
the extremal CGHS black hole the spacetime inside the hole is flat.
There would not seem to be a problem of constructing apparatus that
could remain indefinitely at a fixed position in the horn.} to observe
particles that come down the horn.  At some later time these detectors
could send a signal to the external observer, enabling him to construct
the full quantum S-matrix.

In view of the fact that the CGHS mechanism leads to a scenario which
can resolve all of the paradoxes of black hole evaporation, it is
important to find models in which their mechanism actually works.
Although we have not succeeded in this task in the present paper, we
feel that the possibility of finding a two dimensional Lagrangian for
gravity coupled to one or more scalars that implements the CGHS proposal
is still open.  Of course, even if such a Lagrangian is found, and can
be justified as a well defined approximation to a more realistic theory
of quantum gravity, the CGHS proposal can only deal with very special
situations in which the approximation of a classical geometry, that
varies smoothly on scales much larger than the Planck length, makes
sense.  However, it can serve as a paradigm for a more general
resolution of the \lq\lq Hawking paradox'', which would apply to
regimes which cannot be described by classical differential geometry.
The important point about this mechanism is that it provides a concrete
picture of how black hole evaporation could be compatible with ordinary
quantum mechanics.  We suspect that this picture will survive the
details and difficulties of various models and will form the intuitive
basis for the final understanding of black hole evaporation within the
context of a complete theory of quantum gravity.\foot{We should however
note that even the successful construction of a model exhibiting
the CGHS mechanism does not rule out the
possibility that in some circumstances black hole evaporation does lead
to the creation of a baby universe\dyson .  It would be interesting to
develop an effective theory of such processes, similar to the one we
have constructed for the hypothetical stable configurations proposed by
CGHS.}
\vfill\eject
\centerline{\bf ACKNOWLEDGEMENTS}

\noindent
We would like to thank, E.Martinec, S.Shenker, and particularly
A.Strominger for illuminating conversations.  The work of T.B., A.D.,
and M. O' L. was supported in part by DOE grant DE-FG05-90ER40559 and
that of M.R.D. by an NSF Presidential Young Investigator Grant
PHY-9157016 and an A.P. Sloan Foundation grant BR-3095.

\centerline{\bf NOTE ADDED}
\noindent
After this work was completed we received a preprint by Russo et. al.
(SU-ITP-92-4) in which substantially the same conclusions as ours about
singularities in the CGHS equations were obtained.  We thank L.
Thorlacius for sending us a copy of this manuscript.

\vfill\eject
\centerline{\bf Figure Captions}

\noindent
{\bf Figure 1}: A section of the three dimensional
spatial geometry of the extremal
dilaton black hole with fixed polar angle.

\noindent
{\bf Figure 2}: The Penrose diagram of the radius-time plane of a
hypothetical nonsingular solution of the four dimensional large $N$ equations.
\listrefs
\end